\documentclass[10pt,twocolumn,floatfix,reprint]{revtex4}

\usepackage{graphicx,amssymb,amsmath}

\begin{document}

\title{From mobile phone data to the spatial structure of cities}

\author{Thomas Louail$^{1,2}$, Maxime Lenormand$^3$, Oliva Garc\'{i}a Cant\'u$^4$,
  Miguel Picornell$^4$, Ricardo Herranz$^4$, Enrique Frias-Martinez$^5$, Jos\'e
  J. Ramasco$^3$, Marc Barthelemy$^{1,6}$}

~

\affiliation{$^1$ Institut de Physique Th\'{e}orique, CEA-CNRS (URA 2306), F-91191, 
Gif-sur-Yvette, France}

\affiliation{$^2$G\'eographie-Cit\'es, CNRS-Paris 1-Paris 7 (UMR 8504), 13 rue du
  four, FR-75006 Paris, France}

\affiliation{$^3$IFISC, Instituto de F\'isica Interdisciplinar y Sistemas
Complejos (CSIC-UIB), Campus Universitat de les Illes Balears, E-07122
Palma de Mallorca, Spain}

\affiliation{$^4$Nommon Solutions and Technologies, calle Ca\~nas 8,
  E-28043 Madrid, Spain}

\affiliation{$^5$Telefonica Research, E-28050 Madrid, Spain}

\affiliation{$^6$Centre d'Analyse et de Math\'ematique Sociales, EHESS-CNRS (UMR
8557), 190-198 avenue de France, FR-75013 Paris, France}

\begin{abstract}

Pervasive infrastructures, such as cell phone networks, enable to
capture large amounts of human behavioral data but also provide
information about the structure of cities and their dynamical
properties. In this article, we focus on these last aspects by
studying phone data recorded during 55 days in 31 Spanish
metropolitan areas. We first define an urban dilatation index which
measures how the average distance between individuals evolves during
the day, allowing us to highlight different types of city
structure. We then focus on hotspots, the most crowded places in the
city. We propose a parameter free method to detect them and to test
the robustness of our results. The number of these hotspots scales
sublinearly with the population size, a result in agreement with
previous theoretical arguments and measures on employment datasets. We
study the lifetime of these hotspots and show in particular that the
hierarchy of permanent ones, which constitute the `heart' of the city,
is very stable whatever the size of the city. The spatial structure of
these hotspots is also of interest and allows us to distinguish
different categories of cities, from monocentric and "segregated"
where the spatial distribution is very dependent on land use, to
polycentric where the spatial mixing between land uses is much more
important. These results point towards the possibility of a new,
quantitative classification of cities using high resolution
spatio-temporal data.

\end{abstract}

\maketitle

\section*{Introduction}

Pervasive, geolocalized data generated by individuals has recently
triggered a renewed interest for the study of cities and urban
dynamics, and in particular, individual mobility patterns
\cite{Asgari2013}. Various data sources have been used such as car GPS
\cite{gallotti_towards_2012, Furletti2013}, RFIDs for collective
transportation \cite{roth2011structure}, and also data coming from
social networks such as Twitter\cite{Hawelka2013} or
Foursquare\cite{noulas_tale_2012}. A recent, very important source of
data is given by individual mobile phone data
\cite{onnela2007structure, lambiotte2008geographical}. These data have
allowed to study the individual mobility patterns with a high spatial
and temporal resolution \cite{gonzalez2008understanding,
  schneider2013unravelling, kung_exploring_2013}, the automatic
detection of urban land uses \cite{pei_new_2013}, or the detection of
communities based on human interactions
\cite{sobolevsky_delineating_2013}.

\begin{figure}[!ht]
  \begin{center}
    \includegraphics[width=0.45\textwidth]{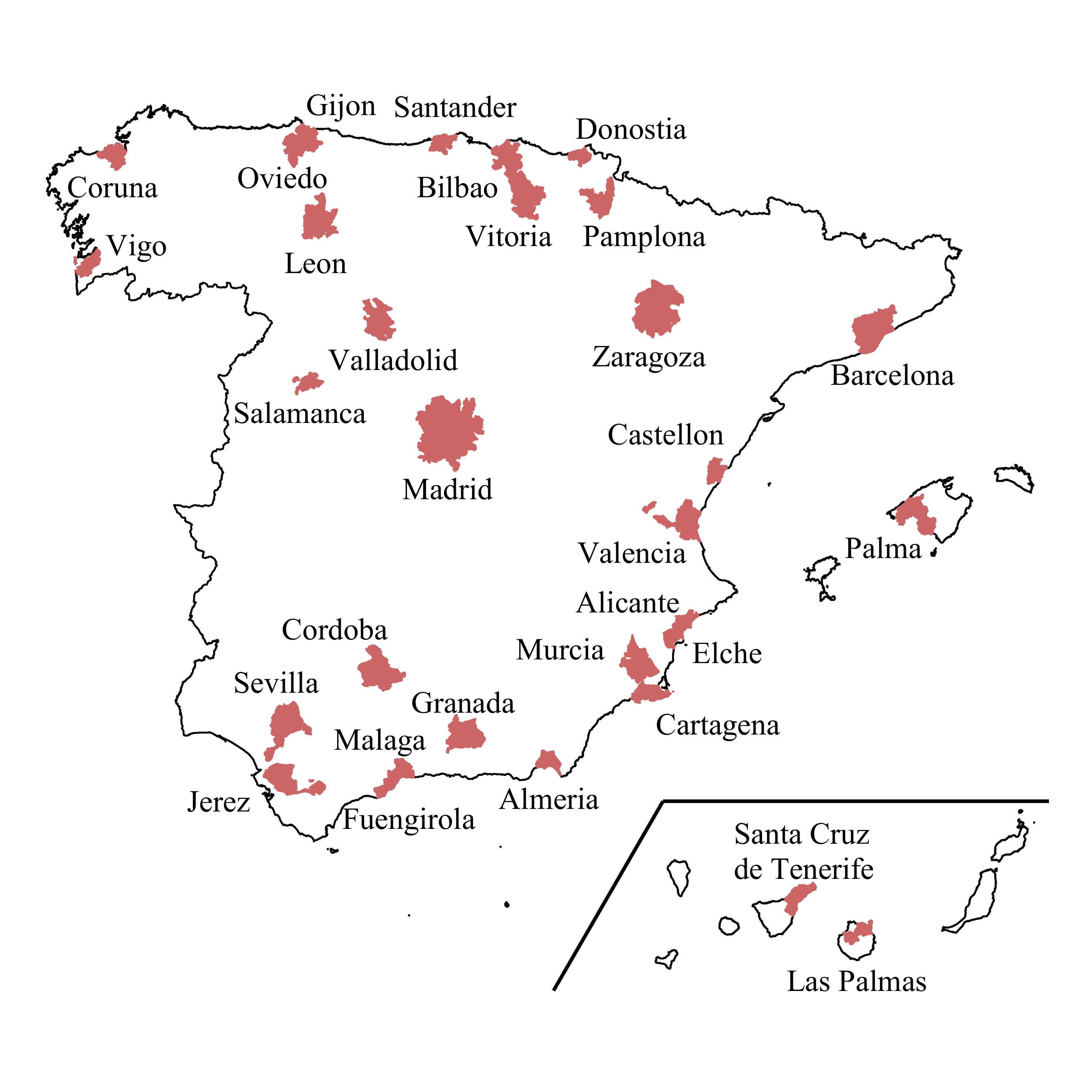}
  \end{center}
  \caption{{\bf The 31 Spanish cities (urban areas) with more
      than 200,000 inhabitants in 2011.} Map of their locations and
    spatial extensions. The set of cities analyzed in this article
    comprises very different types of cities such as central cities,
    port cities and cities on islands.}
  \label{fig:map}
\end{figure}

Morphological aspects, such as the quantitative characterization and
comparison of cities through their density landscape, their space
consumption, their degree of polycentrism, or the clustering degree of
their activity centers, have been studied for a long time in
quantitative geography and spatial economy \cite{anas_urban_1997,
  bertaud2003spatial, Tsai2005, pereira_urban_2013, schwarz2010,
  thomas_morphology_2008, Guerois2008, berroir2011,
  LeNechet2012}. Until the late 2000, these quantitative comparisons
of cities were necessarily based on census data and/or remote sensing
data, both giving a static estimation of the density of individuals
and land uses in the city, at a fine spatial granularity but at a
fixed, unique point in time. Given the atemporal nature of these
studies, they could not investigate some interesting questions related
to the dynamical properties of the spatial structure of cities: how
much does the city shape change through the course of the day?  Where
are the city's hotspots located at different hours of the day?  How
are these hotspots spatially organized? Is there some kind of typical
distance(s) characterizing the permanent core, or `backbone', of each
city? Mobile phone data contain the spatial information about
individuals and how it evolves during the day. These datasets thus
give us the opportunity to answer some of these questions and to
characterize quantitatively the spatial structure of cities. In this
article, we address some of these questions using mobile phone data
for a set of $31$ Spanish cities shown on Figure \ref{fig:map}.
% Phrase à reformuler
We focus on the spatio-temporal properties of cities and, defining new
metrics, study their structural properties and exhibit interesting
patterns of urban systems.

% Results and Discussion can be combined.
\section*{Results}

\begin{figure*}[!ht]
  \begin{center}
    \begin{tabular}{cc}
      (a) & (b) \\
      \includegraphics[width=0.45\linewidth]{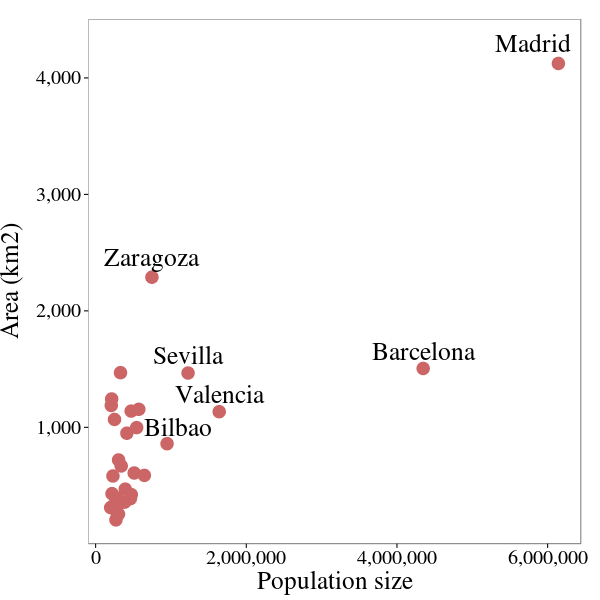}&
      \includegraphics[width=0.45\linewidth]{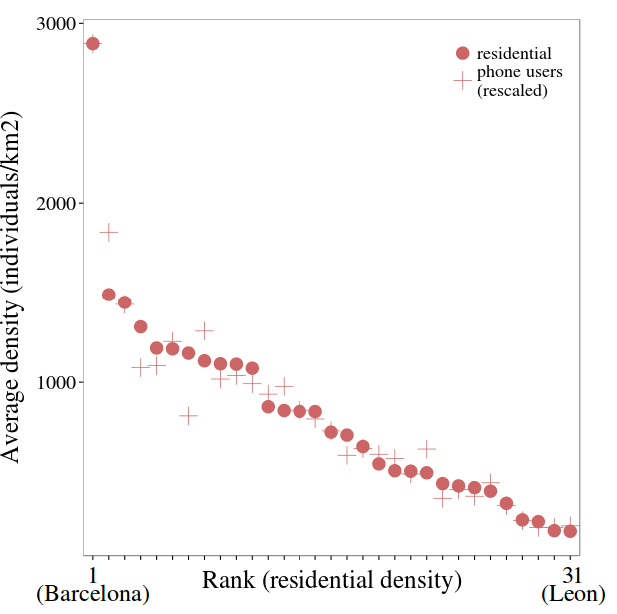}
    \end{tabular}a
  \end{center}
  \caption{{\bf Population sizes, areas and densities of 31 Spanish
      cities (urban areas) with more than 200,000 inhabitants in
      2011.} (a) Population size vs. area. The set of cities under
    study displays a large variety of sizes. We also note that there
    is no general statistical relation between the population size of
    Spanish urban areas and their spatial extension. (b) Rank-size
    distribution of their residential density and phone activity
    density (rescaled by a constant factor given by the inverse of the
    fraction of phone users in the denser urban area, $\rho_{Barcelona, residential} /
    \rho_{Barcelona, phone users}$). The distribution shows that the
    fraction of phone users is almost constant in all cities.}
  \label{fig:dataset}
\end{figure*}

% Our analysis is based on a mobile phone dataset provided by a major Spanish
% phone company and which concerns $31$ Spanish urban areas studied
% during weekdays.
Our analysis is based on aggregated and annonymized mobile phone
traces provided by a Spanish telecommunications operator, which
concerns 31 Spanish urban areas studied during weekdays. These urban
areas are very diverse in terms of geographical location, area,
population size and density, as illustrated by Figure
\ref{fig:dataset}. In particular, the wide range of population sizes
will allow us to test some scaling relations and also to identify
various behaviors. We will first describe the dataset and then present
the results obtained about several aspects of cities.

\subsection*{Data description}

\begin{figure*}[!ht]
  \begin{center}
    \includegraphics[width=\linewidth]{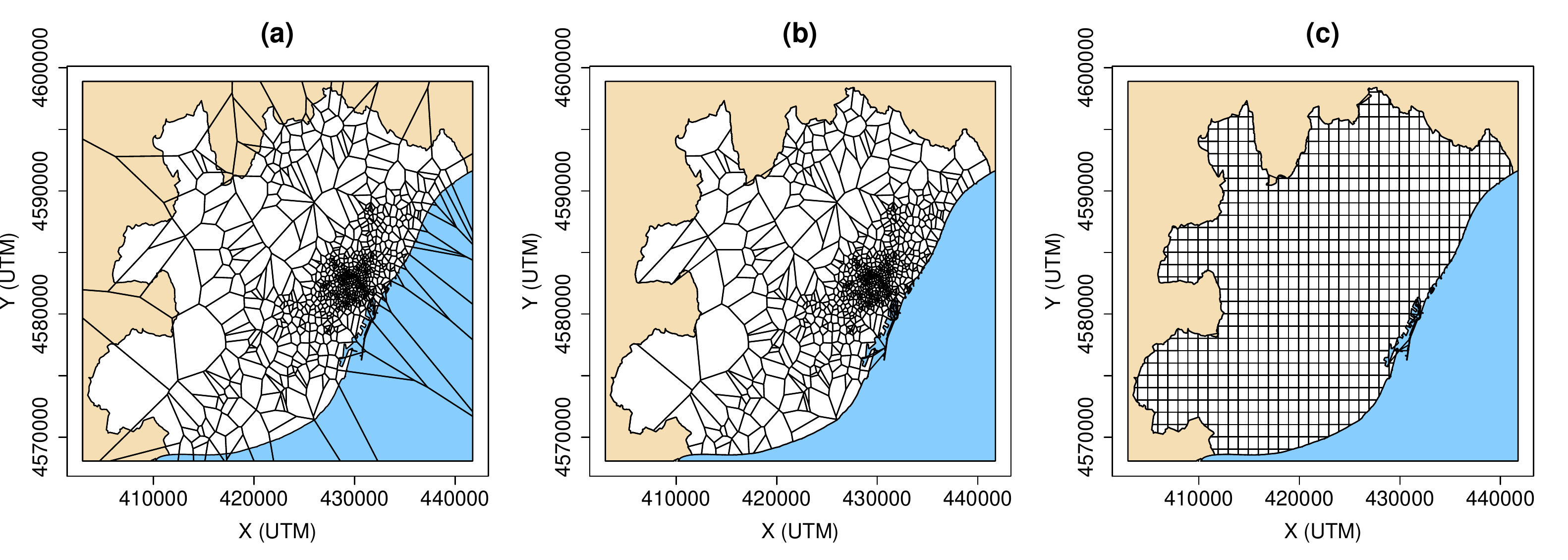}
  \end{center}
  \caption{{\bf Map of the metropolitan area of Barcelona}. The white
    area represents the metropolitan area, the brown area represents
    territories surrounding the metropolitan area and the blue area
    the sea. (a) Voronoi cells of the mobile phone antennas point
    pattern. (b) Intersection between the Voronoi cells and the
    metropolitan area. (c) Grid composed of 1 km$^2$ square cells on
    which are aggregated the number/density of unique phone users
    associated to each phone antenna.}
\label{fig:aggregation}
\end{figure*}

Our analysis is based on a mobile phone dataset provided by a Spanish
telecommunications operator. The aggregated records represent the
number of unique individuals using a given antenna for each hour of
the day. No individual information or records were available for this
study. These data provide some snapshots of the spatial distribution
of people in the city at successive points in time. We have this
information for the 31 Spanish urban areas of more than 200,000
inhabitants, and for 55 days%  (from 1 to 15 September 2009, from 1 to
% 31 October 2009, from 1 to 8 November 2009 and 10 November 2009)
. The number of users (per hour) represents in average $2\%$ of the
total population and at most $5\%$ of the total population. These
percentages are almost the same for all the urban areas. Given the
irregularity of the spatial distribution of the antennas in each city
and from one city to another, we spatially aggregated the
number/densities of users recorded each hour in each mobile phone
antenna on a regular square grid of varying cell size $a$, in order to
simplify comparisons of indicators between cities, as shown on Figure
\ref{fig:aggregation}.  The choice of the spatial scale of data
aggregation is known to be an important source of bias in spatial
analysis \cite{openshaw1979million}, hence we tested the robustness of
our results on three different sizes of grid cells (see section
Methods for details).

\subsubsection*{General features}

\begin{figure*}[!ht]
  \begin{center}
    \includegraphics[width=\linewidth]{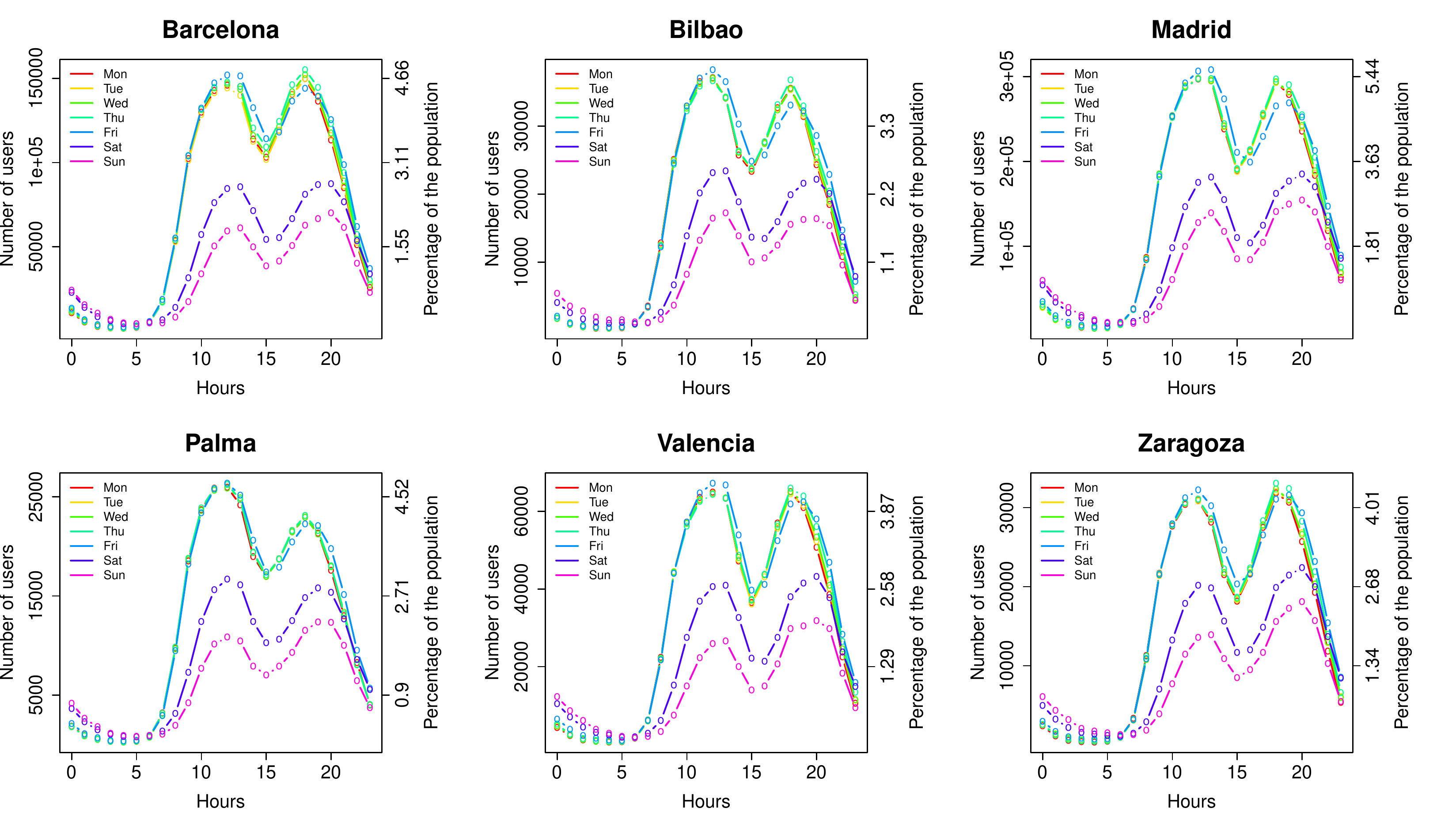}
  \end{center}
  \caption{{\bf Number of mobile phone users according to the hour of
      the day, for each day of the week, in six Spanish
      metropolitan areas.}}
\label{fig:nb-users}
\end{figure*}

\begin{figure*}[!ht]
  \begin{center}
    \begin{tabular}{cc}
      (a) & (b) \\
      \includegraphics[width=.45\linewidth]{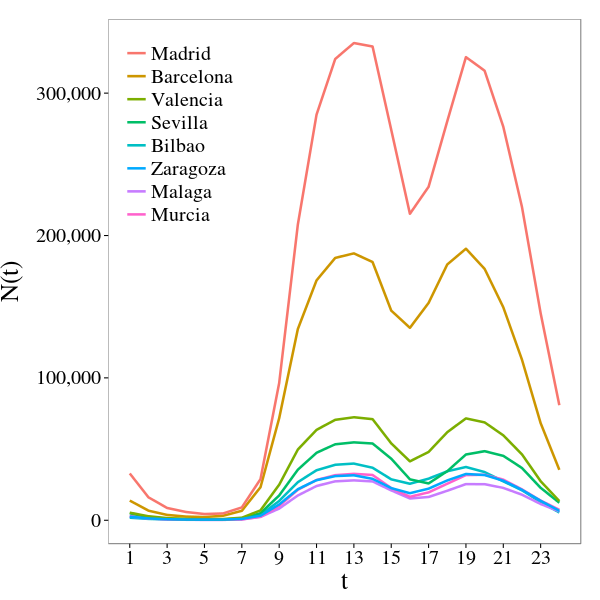}&
      \includegraphics[width=.45\linewidth]{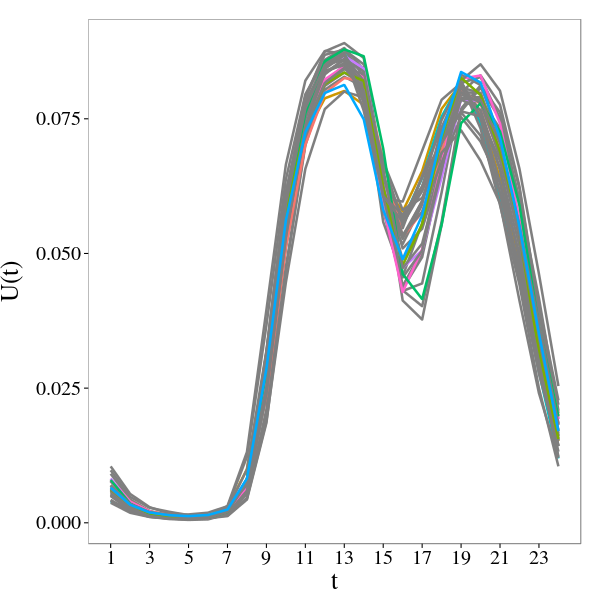}
    \end{tabular}
  \end{center}
  \caption{{\bf Time evolution of the number of mobile phone users per
      hour during an average weekday} (a) Total number of unique
    mobile phone users per hour (shown here for the eight biggest
    Spanish cities). (b) Rescaled numbers of unique users per hour for
    31 cities. Each value $U_i(t)$ is equal to the number
    of phone users in city $i$ at time $t$, $N_i(t)$, divided by the
    total number of phone users in $i$ during the entire day : $U_i(t)
    = N_i(t) / \sum_{t=1}^{t=24}N_i(t)$. The good collapse suggests
    the existence of an urban rhythm common to all cities.}
\label{fig:nb-users2}
\end{figure*}

In order to get a preliminary grasp of the data we plot the time
evolution of the number of users along the day and see if it follows
the same pattern in every city. Figure \ref{fig:nb-users} shows the
average number of mobile phone users per hour according to the day of
the week for six of them. Globally, the number of phone users is
significantly higher during the weekdays than during the weekends,
except at night time. From 11pm to 8am, the number of users is
relatively low, it reaches a minimum at 5am during weekdays and
at 7am during the weekend. For all cities we observe two activity
peaks, one at 12am during weekdays (1pm during the weekend) and
another one at 6pm during weekdays (and at 8pm during the
weekend).
In order to compare these values obtained for different cities, we
rescale the values by the total number of users for an average
weekday. We show the results in Figure~\ref{fig:nb-users2}. The
rescaled plot suggests the existence of a single `urban rhythm' common
to all cities. The data collapse is very good in the morning, while in
the afternoon we observe a little more variability from one city to
another. It is interesting to note that in four cities located in the
western part of Andalusia (Sevilla, Granada, Cordoba and Jerez de la
Frontera) the activity restarts later in the afternoon, around 5pm one
hour later than in the other cities.

\subsubsection*{Global weighted indicators versus hotspots analysis}

Essentially, the mobile phone data give access to the local
density $\rho(i,t)$ of users at a location $i$ and at a time $t$. The
difficulty is then to study this complex object which displays
variation in time and space. We will consider here two
main directions to tackle this problem. The first one is to define
global indicators that consider all points and weight them by the
user density. The second approach consists in identifying local maxima of
the function $\rho(i,t)$, or in other words, the hotspots. There are
pros and cons in each method. Looking at hotspots
is convenient since it provides a clear picture of the important
locations in the city, but contains some arbitrariness in their
determination. On the other hand, working with weighted indices does
not require to identify hotspots but at the cost of producing results
more difficult to interpret.  These two approaches can however be seen
as complementary since they highlight different properties of the
city: weighted indices inform us about the global properties of a
given city, while the hotspots give us a more local look and allow us
to concentrate on the `heart' of the city. This is why
in the following we will successively apply the two methods.

\subsection*{Global analysis}

\subsubsection*{Urban dilatation index}

\begin{figure*}[!ht]
\begin{center}
    \begin{tabular}{cc}
     (a) & (b) \\
     \includegraphics[width=.46\linewidth]{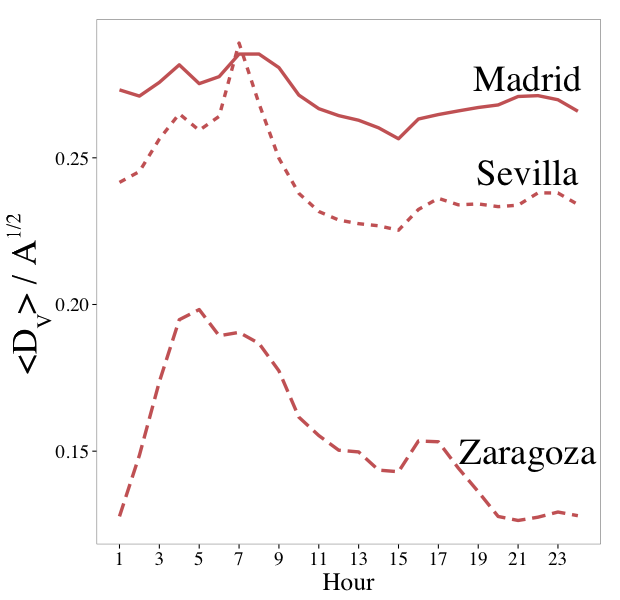}&
     \includegraphics[width=.46\linewidth]{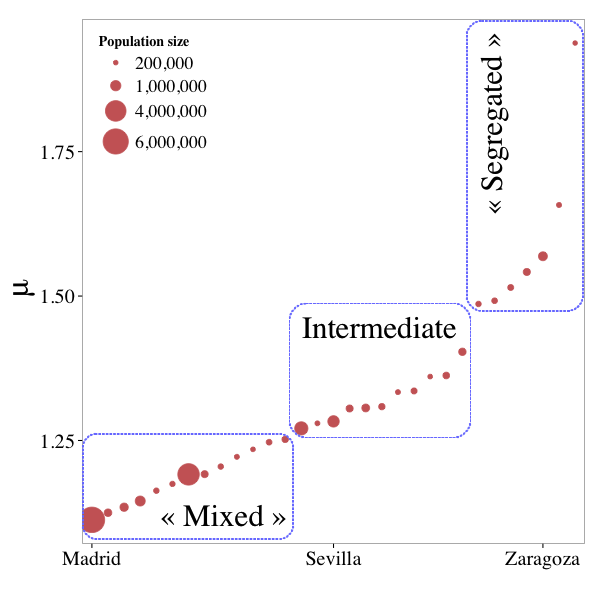}
   \end{tabular}
\end{center}
\caption{{\bf Time evolution of the average distance $D_V(t)$ between
    phone users in the city, and the values of the dilatation index
    $\mu = \max D_V(t)/\min D_V(t)$ for the 31 Spanish metropolitan
    areas studied.} (a) Illustration of the time evolution of $D_V$ in
  three urban areas: Madrid, Sevilla and Zaragoza. This distance $D_V$
  is equal to the average of the distances between each pair of cells
  weighted by the density of each of the cells. The resulting distance
  is then divided by the typical spatial size of the city (given by
  $\sqrt{A}$ the square root of the city's area) in order to compare
  the curves across cities. (b) Rank-size distribution of the
  dilatation index $\mu$ in the 31 metropolitan areas.}
\label{fig:dilatation-index-venables}
\end{figure*}

The average weighted distance $D_V(t)$ between individuals in the
city (see section Methods for the precise definition) and its
evolution during the course of an average weekday provides a first
interesting indicator about the organization of the city. Figure
\ref{fig:dilatation-index-venables} (a) shows the evolution of this
normalized average, weighted distance during a typical weekday. 
We can essentially distinguish two broad categories according to the
spatial organization of residences and activities:
\begin{itemize}
\item In the case of the simple picture of a typical monocentric city
  with predominant Central Business District (CBD), the city collapses
  in the morning when people living in the suburbs commute to their
  workplaces, and expands in the evening when they get back home. We
  then expect in this case a large variation (at the city scale) of
  the average distance $D_V$. In this case, activity and residential
  places are spatially "segregated".
\item For more polycentric cities, where residential and work places
  are spatially less separated, we expect a smaller variation of $D_V$
  than the one observed for monocentric cities. Here activity places
  and residential areas are more "mixed".
\end{itemize}

For all cities we observe the same typical pattern: we see two peaks,
one around 7 am, when people switch on their mobile phones, probably
at home or when they are in transportation system's entry points (see
Figure~\ref{fig:dilatation-index-venables}(a)). We then see a decrease
of the distance (the city `collapses'), displaying spatial concentration
of individuals during the middle of the day, mainly corresponding to
the activity period for most individuals (workers/students). During
the afternoon we see a second, smaller peak dispersed over 4-5pm, when
people start going back home. This afternoon peak is less pronounced,
suggesting a higher variety of mobility behaviors at the end of the
day. The interesting feature of theses curves is the variation
amplitude that informs us about the importance of this collapse
phenomenon. Despite the fact that several factors such as phone use or
behavioral factors affect these variations, we observe a common
pattern: a pronounced peak at the beginning of the day and a minimum
usually observed at the middle of the day. From this curve it is then
natural to calculate for each city a `dilatation coefficient' defined
as
\begin{equation}
\mu=\frac{\max_t(D_V (t))}{\min_t(D_V(t))}
\end{equation}
We show in Figure~\ref{fig:dilatation-index-venables}(b) the rank plot
of this dilatation index obtained for the 31 cities where we can
distinguish roughly three groups of cities. For the first group with a
value of $\mu$ around one, the average distance stays approximately
constant throughout the day. This means that whatever the hour of the
day, the spatial spread of the high density locations does not change
significantly. High density locations correspond to different
activities depending on the moment of the day, and a small value of the
dilatation coefficient implies that daytime activity places (work
places, schools, leisure places) are not more spatially concentrated
than residences. Homes and activity places are more entangled,
supporting the picture of more `mixed' cities, such as Madrid for
example. In the opposite case of large values, the spatial
organization of the different high-density locations changes
significantly along the day. A typical example would be a monocentric
city where individuals are localized in the CBD during the day and
where residences are spread all around the center. In our set, Zaragoza
for example is representative of this type of cities. For the
intermediate group the cities display a less marked behavior, probably
resulting of a combination of monocentric and polycentric features.
% The dilatation index value does not seem corellated to the size of
% the city, we find some of the biggest cities in the three groups

\subsection*{Hotspots analysis}

\subsubsection*{Identifying the hotspots} 

This problem corresponds to identify local maxima in the surface of
density of users. A simple method amounts to choose a threshold
$\delta$ and to consider that every point $i$ with a density larger
than this threshold $\rho(i,t)>\delta$ is a hotspot at time $t$. Most
of the methods so far rely on this simple argument but there is
obviously some arbitrariness in the choice of $\delta$. In contrast
here (all technical details can be found in the Methods section), we
discuss two extreme choices for the threshold value. The lower
threshold $\delta_{min}$ corresponds to the average value of the
density, which is indeed a reasonable, minimal requirement to be a
local maxima. Based on considerations about the Lorenz curve of the
density, we are also able to determine another value $\delta_{max}$
which can be considered as the maximal, reasonable value for
$\delta$. In the following we will distinguish the `Average' method
from the `Loubar' method which correspond to the two values
$\delta_{min}$ and $\delta_{max}$, respectively. The most important
point here, is that once the lower and upper bounds for the threshold
are determined and allow for the identification of hotspots, all the
results obtained should be robust with respect to the choice of
$\delta$. In other words, if a given result is qualitatively the same
when considering the lower and upper bounds for $\delta$, the result
can safely be considered as an intrinsic feature of the system.

\subsubsection*{Number of hotspots} 

\begin{figure}[!ht]
\begin{center}
  \includegraphics[width=\linewidth]{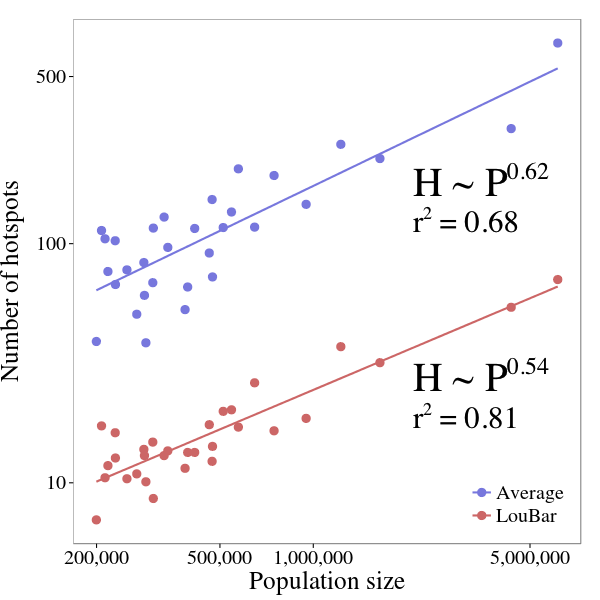}
\end{center}
\caption{{\bf Scatter plot and fit of the number of hotspots $H$
    vs. the population size $P$ for the 31 cities studied.} Each point
  in the scatterplot corresponds to the average number of hotspots
  determined for each one-hour time bin of a weekday (for five
  weekdays considered here). The power law fit is consistent, for both
  hotspots identification methods, with a sublinear behavior
  characterized by an exponent of order $0.6$, a value in agreement
  with theoretical predictions and empirical observations on
  employment data \cite{LoufBarthelemy2013}.}
\label{fig:HvsP}
\end{figure}

We first focus on the number of hotspots. Using both methods,
'Average' and `Loubar', for each city we count the number of hotspots
at each hour of the day, compute the average over the day and see how
this average number scales with the population size of the city. This
measure is motivated by recent theoretical and empirical work
\cite{LoufBarthelemy2013} that has highlighted a clear sub-linear
relation between the population size of cities and their number of
activity centers (defined as employment hotspots). For the U.S., it
has been shown that the number of activity centers $N_a$ (determined
from employment data) scales as
\begin{equation}
N_a\sim P^{\beta}
\end{equation}
with $\beta\sim 0.64$. Figure \ref{fig:HvsP} displays the number $H$
of hotspots versus the population for the set of the 31 biggest
Spanish cities considered here. The power law fit confirms the result
obtained in \cite{LoufBarthelemy2013} that there is a sublinear
relation and, remarkably enough, that the value of the exponent is of
the same order. We note here that this result is robust against the
thresholding criteria used to define hotspots (see also section Methods for
aggregation grids with different cell sizes).
We also note here that recent empirical work \cite{arcaute_city_2013}
has highlighted the sensitivity of the values of scaling laws
exponents to the choice of city boundaries. This result underlines the
crucial role of city definition when attempting to identify patterns
of behavior across cities, and the need for consistency in defining
the spatial boundaries of cities for such comparisons
\cite{bretagnolle_organisation_2009}. That is the reason that has led us
to rely on the spatial delimitations of \emph{urban areas}, which are
harmonized delimitations based on the ratio of home-work commuting
journeys (see Methods for details).

\subsubsection*{Stability of the hotspots hierarchy} 

\begin{figure*}[!ht]
\begin{center}
  \includegraphics[width=\linewidth]{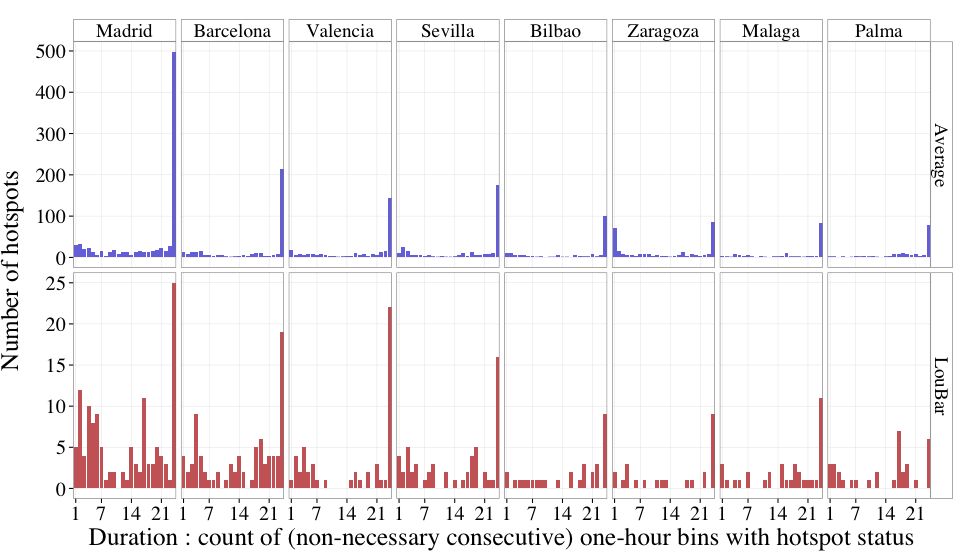}
\end{center}
\caption{{\bf Histogram of lifetime duration of hotstpots for eight cities
  and for the two hotspots identification methods (top: `Average'
  method and bottom: `Loubar' method).} In the case of the 'Loubar'
  hotspots, we can essentially distinguish three groups: the permanent
  (24h hotspots), intermittent (from 1 up to 7 hours) and
  intermediary (all the others) hotspots.}
\label{fig:persistence}
\end{figure*}

Another interesting feature to inspect in cities is the stability of
their hotspots and the evolution of their relative importance in the
city according to the hour of the day, which is related to the
evolution of the hierarchy of places in the city. In order to capture
the behavior of cities about these aspects, we plot various
indicators. We start with the histogram of the persistence of
hotspots: for each city we count the number of one-hour time bins
during which each cell has been a hotspot. We then plot the
distribution of the hotspots `lifetime' (measured in number of one-hour
bins), as shown in Figure~\ref{fig:persistence} for the eight largest
Spanish cities.
Figure~\ref{fig:persistence} highlights the importance of `permanent'
hotspots, i.e. locations which are hotspots during the whole day. Each
city has its number of important locations, those that form the
`heart' of the city. In addition to the permanent hotspots we also
observe two other main groups: a set of intermediate hotspots
(with lifetime of the order half a day) and `intermittent' hotspots
that are present only a few hours per day. We note that these groups
are robust with respect to the hotspot definition, that is when
defined with the `Average' criterion (top line of each histogram) and
with the `LouBar' criterion (bottom line).

The permanent hotspots are the most important locations in the city in
terms of individuals density. An interesting question is whether their
rank (according to the density) is constant or changes during the
day. In order to test the stability in time of the hierarchy of
permanent hotspots, we compute the Kendall tau value $\tau(t)$ of the
set of permanent hotspots (see the Methods section for definition and
for the plots). Our results show that the heart of the cities is
indeed very stable both in space and in time, whatever their size.

\subsubsection*{Spatial structure of the hotspots}

\begin{figure*}[!ht]
\begin{center}
    \begin{tabular}{cc}
     (a) & (b) \\
     \includegraphics[width=.5\linewidth]{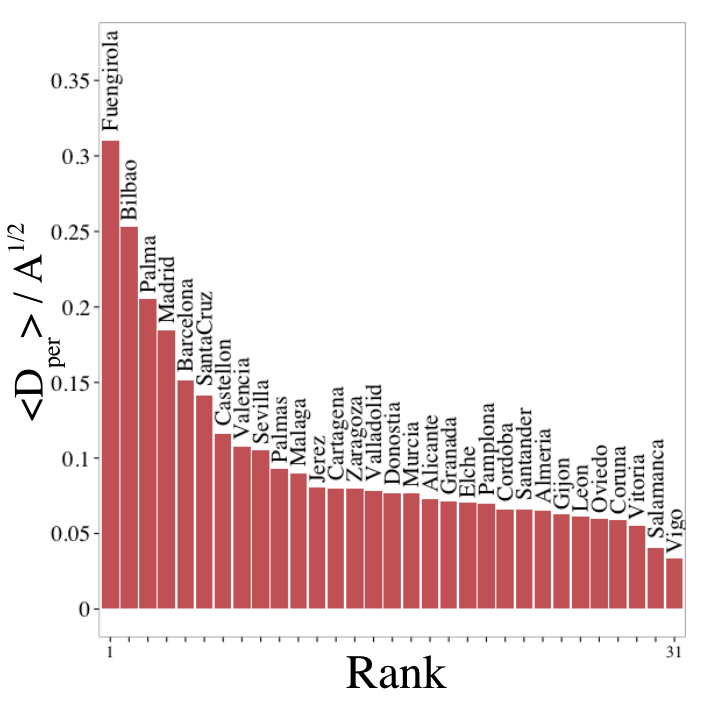}&
     \includegraphics[width=.5\linewidth]{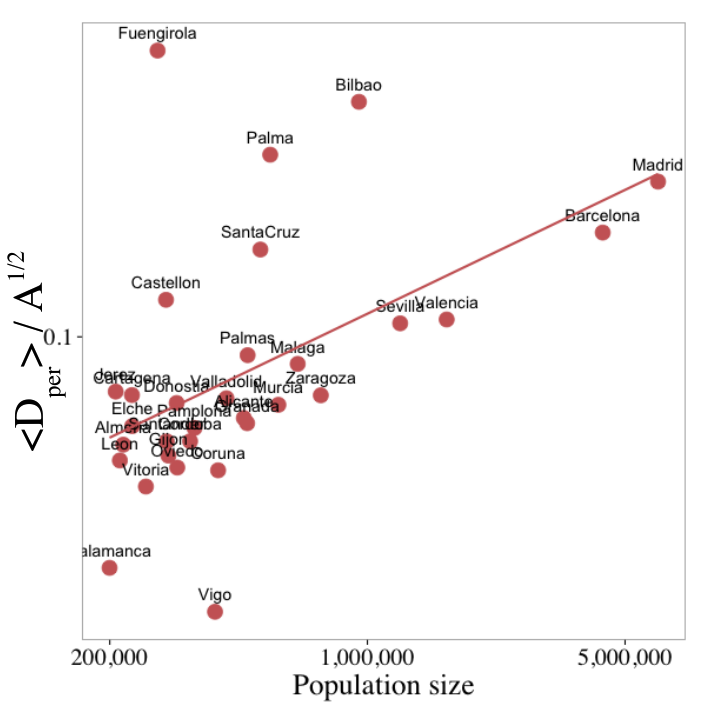}\\
      (c) & (d)\\
      \includegraphics[width=.5\linewidth]{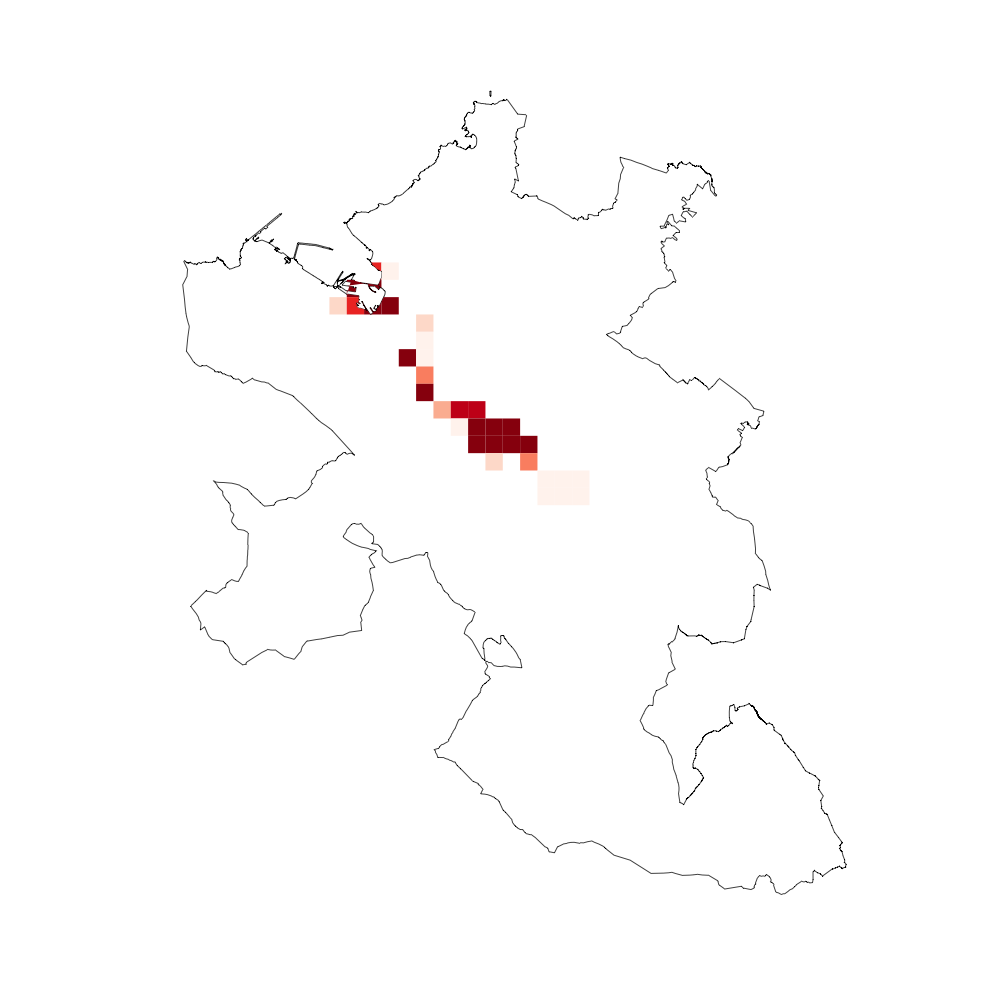}&
      \includegraphics[width=.5\linewidth]{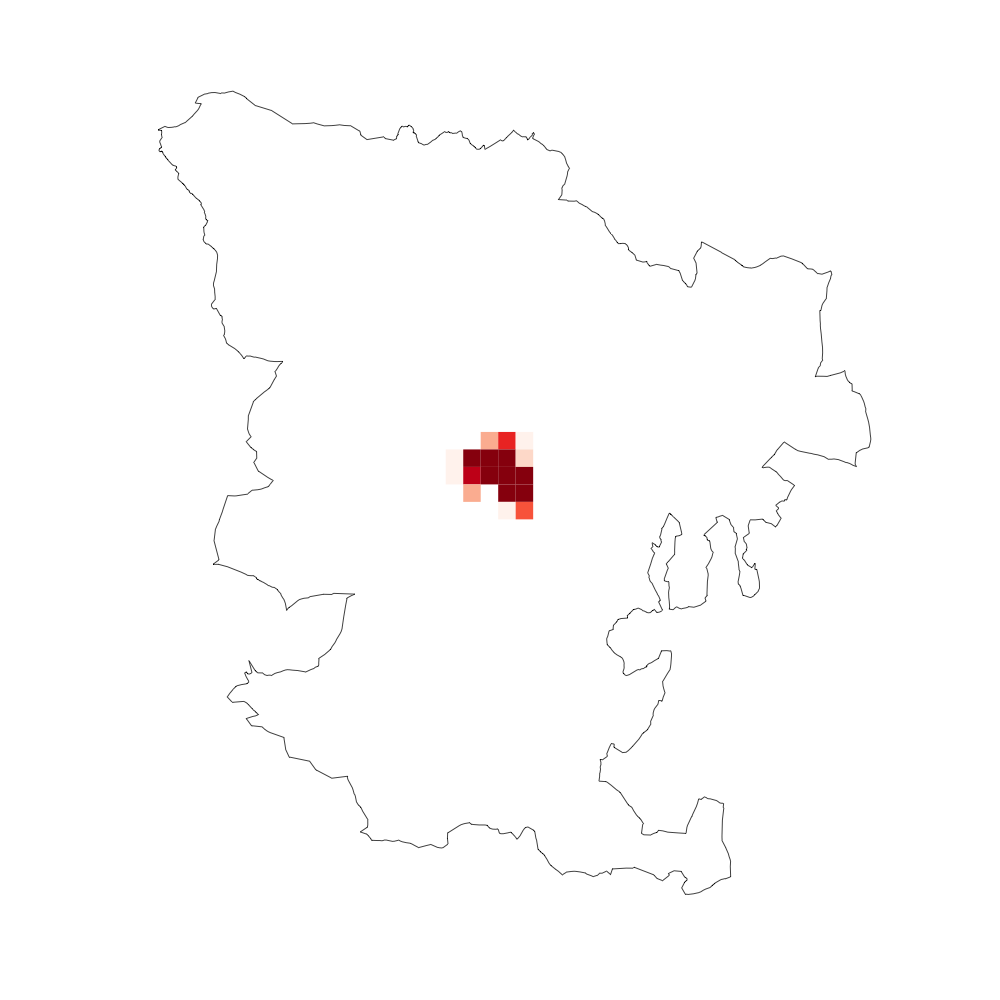}
   \end{tabular}
  \end{center}
  \caption{{\bf Different spatial structure of hotspots in cities.}
    Rank plot of the compacity coefficient $C=\langle D_{per}\rangle /
    \sqrt{A}$ among the 31 metropolitan areas. (b) Compacity versus
    population size. We observe a trend (at least for a large subset
    of cities, the corresponding fit is shown as a guide to the
    eye). (c) and (d) The spatial organization of the $1 km2$
    permanent hotspots determined by the Loubar method, in the urban
    areas of Bilbao ($950,000$ inhabitants) and Vitoria ($250,000$
    inhabitants). These figures reveal two types of spatial
    organization: polycentric in the case of Bilbao (c), whose
    permanent hotspots are not contiguous and more spread over the
    space of the urban area, and clearly compact and monocentric in
    the case of Vitoria (d).}
  \label{fig:core-compacity}
\end{figure*}

Another important question about hotspots concerns their spatial
organization. We start with the specific group formed by the permanent
hotspots, as defined by our more restrictive criteria `LouBar' (see
Methods section). We compute how distant they are from each other,
compared to the typical size of the city given by $\sqrt{A}$, where $A$ is the
city's area. We show in Figure \ref{fig:core-compacity} the rank-plot of
our `compacity coefficient' defined as
\begin{equation}
C(i)=\frac{\langle D_{per(i)}\rangle}{\sqrt{A_i}}
\end{equation}
where $\langle D_{per(i)}\rangle$ is the average distance between
permanent, weekday hotspots in city $i$, and $A_i$ is the area of the city
$i$. This indicator informs us how the permanent hotspots are sprawled
all over the city's space, and it is thus a measure of the compacity of
the city core: for cities with values around 0, the permanent hotspots
are very close to one another, when compared to the spatial extension
of the urban area. On the contrary, a value close to one indicates
that these always-crowded places are spread all over the whole city
space (see figure \ref{fig:core-compacity}).
It is interesting to note in Figure~\ref{fig:core-compacity}(b) that the
compacity of a city seems to increase with the population size. At
least for a large subset of cities, we indeed observe this trend,
which is consistent with the idea that the larger the city, the more
spread are the hotspots (and the more polycentric it tends to be).

\begin{figure}[!ht]
  \begin{center}
    \begin{tabular}{cc}
      (a) & (b) \\
      \includegraphics[width=.5\linewidth]{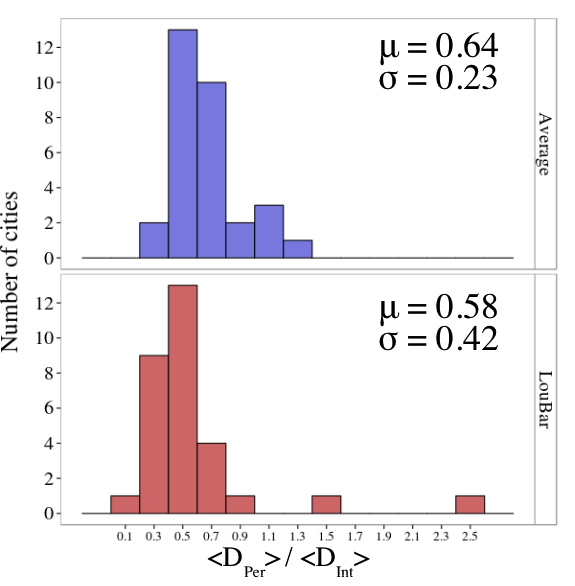}&
      \includegraphics[width=.5\linewidth]{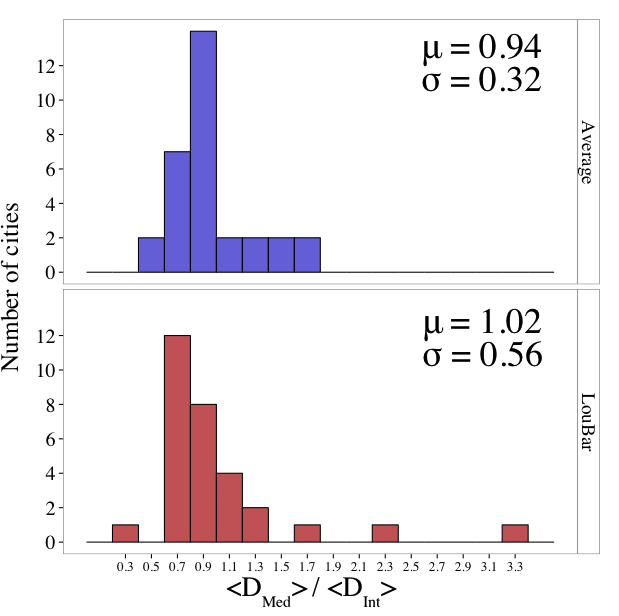}
    \end{tabular}
  \end{center}
  \caption{{\bf Histograms of the coefficients $<D_{per}> / <D_{int}>$ (a)
      and $<D_{med}> / <D_{int}>$ (b).} While the spatial features of
    intermittent and intermediary hotspots are similar, the main
    difference between cities lies in how the permanent hotspots are
    distributed in space.}
  \label{fig:HS-spatial-structure}
\end{figure}

For each city, once we have determined the hotspots and have
classified them into permanent, intermediary and intermittent, we
measure the average distance between hotspots within each group. For
example we can look at $\langle D_{int~hotspots}\rangle/\langle
D_{per~hotspots}\rangle$, the ratio between the typical distance
separating intermittent hotspots and the typical distance separating
permanent hotspots. Since the intermittent hotspots are those with a
lifespan of six hours at most, they are more inclined to capture the
residential locations, while the permanent hotspots represent the
dominant places of the city, that is, zones that are very dense both
during daytime and nightime. On Figure \ref{fig:HS-spatial-structure}
(a) we plot the histogram of this ratio for all cities, for the two
hotspots delimitation criteria (see section Methods for these plots
with different sizes of the aggregation grid). We can see in this plot
that the distribution is centered around 0.6 (with similar results for
the more restrictive Loubar criterion).  We also computed the ratio of
the average distance between intermittent hotspots and the average
distance between intermediary hotspots (i.e. those that are not
intermittent or permanent, so those who are present between 7 and 23
hours per day). We plot the histogram of this ratio for all 31 cities
in Figure \ref{fig:HS-spatial-structure} (b). The distribution is
peaked around 0.95-1, with lower values of standard deviation, which
means that intermittent and intermediate hotspots are, on average, as
much dispersed and that the significative differences lie in the
spatial organization of permanent hotspots vs. non permanent hotspots.

\section*{Discussion}
\label{sec:discussion}

We have shown in this study that it is possible to extract relevant
information from mobile phone data, not only about the mobility
behavior of individuals, but also about the structure of the city
itself. We have defined various indices that allow us to propose a new
classification of cities based on their dynamical properties. We have
also presented a method to determine the dominant centers, the
hotspots, and we have confirmed recent results -obtained on completely
different data- showing that the number of activity centers in cities
scales sublinearly with the population size of the city. We have also
highlighted some properties of hotspots such as the strong stability
of the hierarchy of city centers along the day, whatever the city
size. These results constitute a step towards a quantitative typology
of cities and their spatial structure, an important ingredient in the
construction of a science of cities.

% You may title this section "Methods" or "Models". 
% "Models" is not a valid title for PLoS ONE authors. However, PLoS ONE
% authors may use "Analysis"

\section*{Methods}
\label{sec:methods}

\subsection*{Spatial delimitation of cities}

Comparing the spatial structure of cities of very different population
sizes and areas requires to rely on a harmonized definition of cities
that goes beyond the arbitrariness of the spatial boundaries of the
administrative units \cite{bretagnolle_time_2002,
  bretagnolle_organisation_2009}. To that extent, we have chosen to rely on
the \emph{urban areas} defined by the AUDES initiative (Areas Urbanas De
ESpa\~na)\footnote{Documentation and open data available at
  http://alarcos.esi.uclm.es/per/fruiz/audes/} which capture
some coherent delimitations of cities regarding the home-work
commuting patterns of individuals living in the core city of the
metropolitan areas and in their surrounding municipalities. These
delimitations are built upon statistical criteria based on the
proportion of residents of surrounding municipalities that commute to
the main city to work.

\subsection*{Average distance between individuals and dilatation
  index}

We started with the Venables index \cite{pereira_urban_2013}, defined as:

\begin{equation}
V=\sum_{i\neq j}s_i s_j d_{i,j}
\end{equation}

with $s_i (t)=n_i (t)/N(t)$ the share of individuals present in cell
$i$ at time $t$, and $d_{ij}$ the distance between $i$ and $j$. When
all activity is concentrated in one spatial unit only, the minimum
value zero of $V$ is reached. An important point of this dilatation
index is that one doesn't need to determine hotspots to compute it. By
normalizing $V$ by the densities, we can compute a weighted average
distance, the `Venables distance'

\begin{equation}
D_V(t)= \frac{\sum_{i<j}s_i(t)s_j(t)d_{ij}}{\sum_{i<j}s_i(t)s_j(t)}
\end{equation}

with $s_i(t)=n_i (t)/N(t)$ the share of individuals present in cell
$i$ at time $t$. In order to compare the value of $D_V$ across cities,
we compute $D_V (t) / \sqrt{A}$ with $A$ the area of the city. By
considering all pairs of cells and weighting their distance by the
densities of individuals in each of them, $D_V(t)$ signals how much the
important places of the city at time $t$ are distant from each other.

\subsection*{Identification of the hotspots}
\label{sec:method-hotspots}

The data gives access to the spatial density $\rho(i,t)$ of users at
different moments. The full density is a complex object and we have to
extract relevant and useful information. The locations that display a
density much larger than the others - the hotspots - give a good
picture of the city by showing where most of the people are. The
hotspots thus contain important information about points of interest
and activities in the city.

% Hoover and LouBar delimitation methods

The determination of centres and subcentres is a problem which has
been broadly tackled in urban economics \cite{giuliano1991subcenters,
  mcmillen2001nonparametric, mcmillen2003number}. Starting from a
spatial distribution of densities, we have to identify the local
maxima. This is in principle a simple problem solved by the choice of
a threshold $\delta$ for the density $\rho$: a cell $i$ is a hotspot
at time $t$ if the instantaneous density of users $\rho(i,t)
>\delta$. This is for example what was done in
\cite{giuliano1991subcenters} to determine employment centres in Los
Angeles. It is however clear that this method introduces some
arbitrariness due to the choice of $\delta$, and also requires prior
knowledge of the city to which it is applied to choose a relevant
value of $\delta$. Nonparametric methods have also been applied to
determine the number of centres, some based on the regression of the
natural logarithm of employment density on distance from the centre
\cite{mcmillen2001nonparametric}, some on the exponent of the negative
exponential fit of the density distribution
\cite{griffith1981modelling}. Limits of these methods stand in the
fact that they return a unique number of centres that could be biased
when the actual density distribution is not properly fitted by an
exponential law. Here we will propose an alternative method that
allows us to control the impact of this choice.

A first simple criterion is to choose the point that corresponds to
the average $m(t)=\overline{\rho(i,t)}$ of the distribution at time
$t$ : all the cells whose density is larger than $m$ are
hotspots. This is indeed a weak definition of what can be considered
as a hotspot, and we propose here to use it as a `lower' bound
$\delta_{min}=m$.

In order to understand how the various properties of hotspots will
depend on this definition, we introduce a more restrictive definition
which will be considered as an upper bound of what can be considered
as a hotspot. In the following we discuss how to find this upper
bound. In order to characterize the disparity of the activity in the
city and to isolate the dominant places, we first plot the Lorenz
curve of the density distribution in the city at each hour. The Lorenz
curve, a standard object in economics, is a graphical representation
of the cumulative distribution function of an empirical probability
distribution. For a given hour, we have the distribution of densities
$\rho(i,t)$ and we sort them in increasing rank, and denote them by
$\rho(1,t)<\rho(2,t)<...<\rho(n,t)$ where $n$ is the number of cells.
The Lorenz curve is constructed by plotting on the x-axis the
proportion of cells $F=i/n$ and on the y-axis the corresponding
proportion of users density $L$ with:
\begin{equation}
L(i,t)=\frac{\sum_{j=1}^i\rho(j,t)}{\sum_{j=1}^n\rho(j,t)}
\end{equation}
If all the densities were of the same order the Lorenz curve would be
the diagonal from $(0,0)$ to $(1,1)$. In general we observe a
concave curve with a more or less strong curvature, and the area between
the diagonal and the actual curve is related to the Gini coefficient,
an important indicator of inequality used in economics.

\begin{figure}[!ht]
  \centering
  \includegraphics[width=\linewidth]{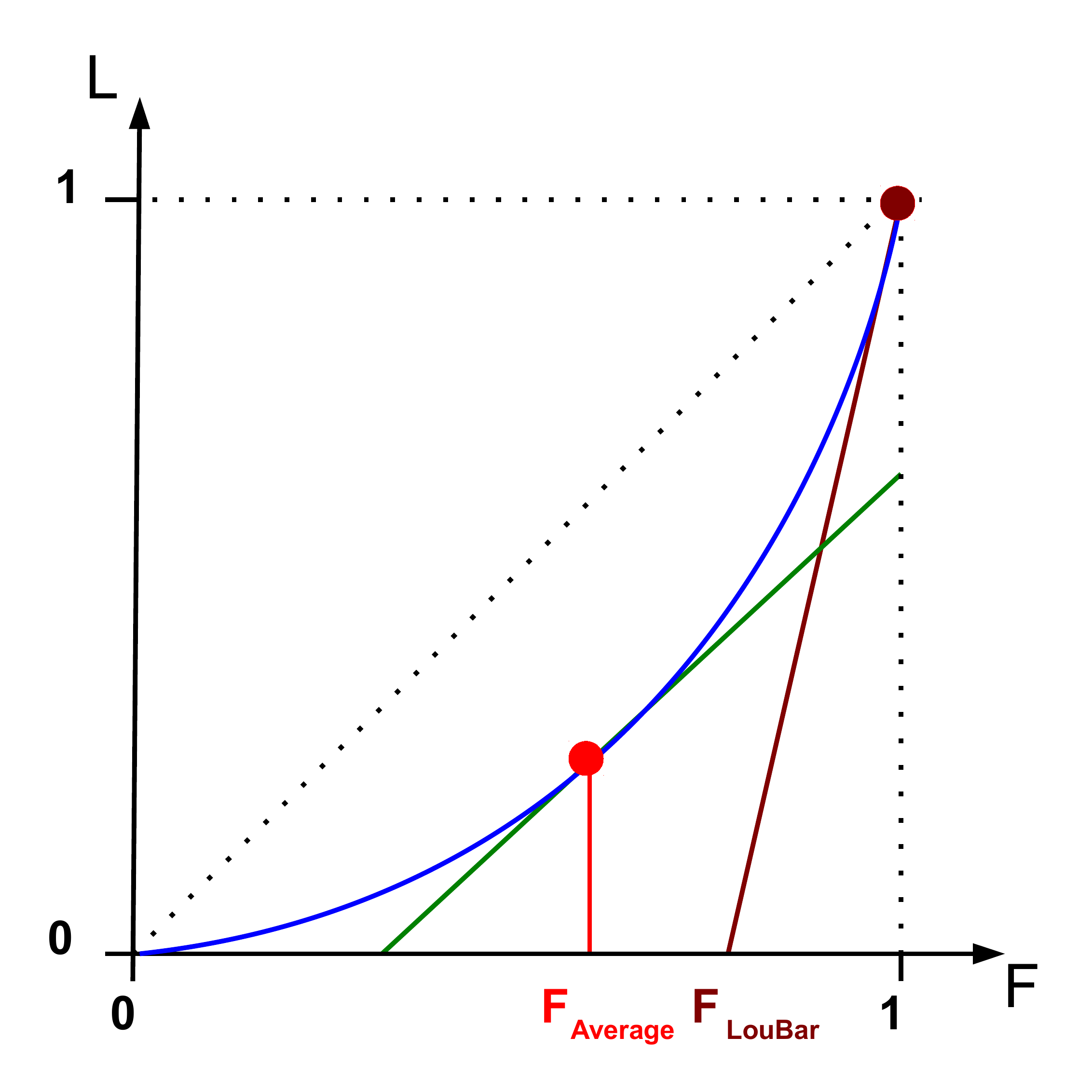}
  \caption{{\bf Illustration of the criteria selection on the Lorenz curve.}}
  \label{fig:Lorenz-curve}
\end{figure}

In the Lorenz curve, the stronger the curvature the stronger the
inequality and, intuitively, the smaller the number of hotspots. This
remark allows us to construct a new criterion by relating the number
of dominant hotspots (i.e. those that have a very high value compared
to the other cells) to the slope of the Lorenz curve at point $F=1$:
the larger the slope, the smaller the number of dominant individuals
in the statistical distribution. The natural way to identify the
typical scale of the number of hotspots is to take the intersection
point $F^*$ between the tangent of $L(F)$ at point $F=1$ and the
horizontal axis $L=0$ (see Figure \ref{fig:Lorenz-curve}). This method
is inspired from the classical scale determination for an exponential
decay: if the decay from $F=1$ were an exponential of the form
$\exp{-(1-F)/a}$ where $a$ is the typical scale we want to extract,
this method would give $1-F^*=a$.
We note here that the average criterion corresponds to the point of
the Lorenz curve with slope equal to $1$. Indeed, the general
expression of the Lorenz curve for the set of densities $\rho(i,t)$
whose cumulative function is $F(\rho)$ is:
\begin{equation}
L(F)=\frac{1}{m}\int_0^F\rho(F)dF
\end{equation}
where $\rho(F)$ is the inverse
function of the cumulative. This point thus satisfies
\begin{equation}
\frac{dL}{dF}=1
\end{equation}
which gives $m=\rho(F_{Avg})$ or in other words, the hotspots will be
those with densities larger than the average. In contrast, our more
restrictive
criterion based on the slope at $F=1$ gives 
\begin{equation}
F^*=1-\frac{\mu}{\rho_M}
\end{equation}
where $\rho_M$ is the maximum value of $\rho(i,t)$ (for a given time
$t$). We thus see that in general $F_{Avg}<F^*$ and that this new
criterion, more restrictive, does not only depend on the average value
of the density but also on the dispersion: as $\rho_M$ increases, the
value of $F^*$ increases and therefore the number of detected hotspots
decreases.

\begin{figure}[!ht]
  \centering
  \includegraphics[width=\linewidth]{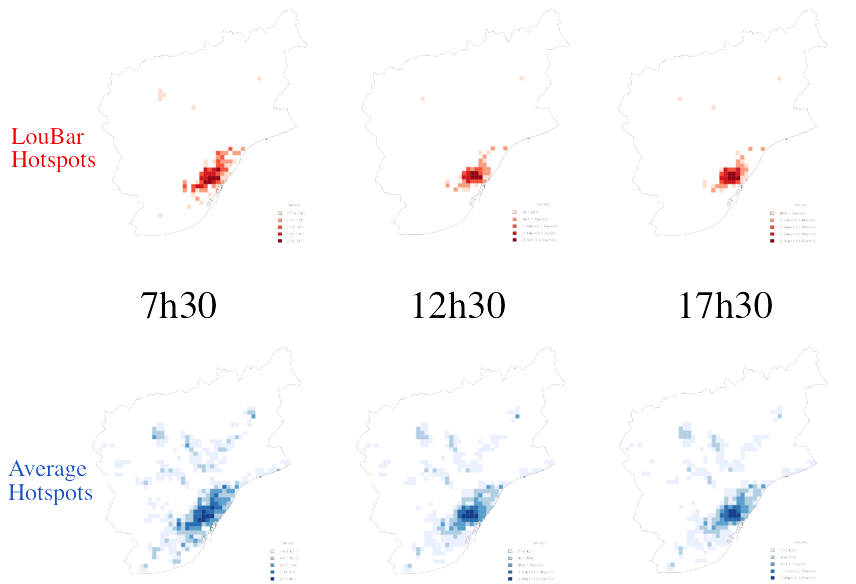}
  \caption{{\bf Location of the hotspots in the metropolitan area of
      Barcelona, selected with two different criteria: the Average
      criterion and our more restrictive criterion ('LouBar').} Here
    density data are aggregated on a grid composed of $1km^2$ square
    cells.}
  \label{fig:mapsHS-methods-BCN}
\end{figure}

All other possible and reasonable methods will then give a value
comprised in the interval $[F_{Avg},F^*]$ between the average
criterion and our criterion (also denoted by `LouBar'). Instead of
choosing a particular point, we will thus study most of the properties
computed for hotspots with the two methods, giving us both a lower and
upper bounds. In particular, we will be able to test the robustness of
our results against the arbitrariness of the hotspot identification
method. Figure~\ref{fig:mapsHS-methods-BCN} shows the location of the
hotspots selected according to the two methods/criteria at different
moments of the day, in the metropolitan area of Barcelona. These maps
can be regarded as the extremes of hotspots maps that reasonable
delimitation methods could produce (i.e. with a number of hotspots
comprised between $F_{Avg}$ and $F^*$.

\subsection*{Influence of the spatial scale of aggregation}
\label{sec:method-aggregation}

\subsubsection*{Hotspots}

\begin{figure}[!ht]
  \centering
  \includegraphics[width=\linewidth]{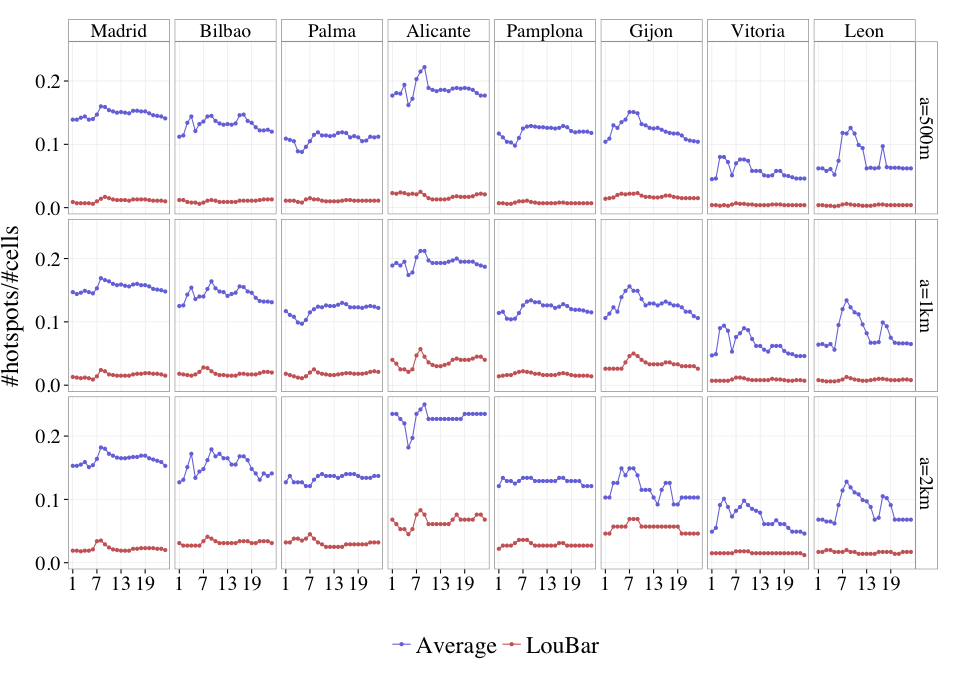}
  \caption{{\bf Time evolution of the ratio
      $\frac{\#hotspots}{\#cells}$ for two hotspots definitions and
      different sizes of grid cells, for eight different cities of
      very different sizes}. The cities chosen cover the full range of
    the poulation size distribtion of the set of the 31 cities
    studied. Every reasonable method for defining hotspots would give
    a value between the two lines of each plot. One can see that
    qualitatively pattern stays identical whatever the grid size for
    couple (city, method).}
  \label{fig:HoverNvsA-cities}
\end{figure}

In the hotspots identification process, the size of the grid cells on
which we aggregate the numbers/densities of users is another arbitrary
parameter (cf. section Methods). Since we don't want to determine this
value separately for each city, we consider that several sizes should
be tested for each city and that it is reasonable to consider that
this cell size $a$ can vary from $500$ meters to $2$ km. Figure
\ref{fig:HoverNvsA-cities} gives an idea of how much the proportion of
hotspots change from one cell size to another. The cell size $a$
should primarily be chosen based on what is considered as a reasonable
size for an urban hotspot. From the pedestrian point of view, every
size between 500 metres and 2 kilometres seems \emph{a priori}
acceptable. Below 500m, it would clearly be necessary to aggregate
contiguous hotspots : for example, for $a=100m$ ($10^{-2} km^2$
cells), two contiguous hotspots could not as easily be distinguished
as two different ones from a pedestrian point of view. In contrast, a
size of $2000m$ can be considered as an upper bound for the same
reasons: if two contiguous cells are classified as hotspots, it is
reasonable to identify them as two distincts neighbourhoods. It is
however a question of perception and should be discussed carefully. In
the hypothesis of $a=1000m$ ($1 km^2$ cells), we chose to consider
that \emph{two adjacent hotspots are two different hotspots}. For
reasonable sizes of grid, the values of the indicators should be
robust with a change of the cell size. We then tested the sensitivity
of our results with respect to different resolutions.

\subsubsection*{Number of hotspots}

\begin{figure}[!ht]
\begin{center}
  \includegraphics[width=\linewidth]{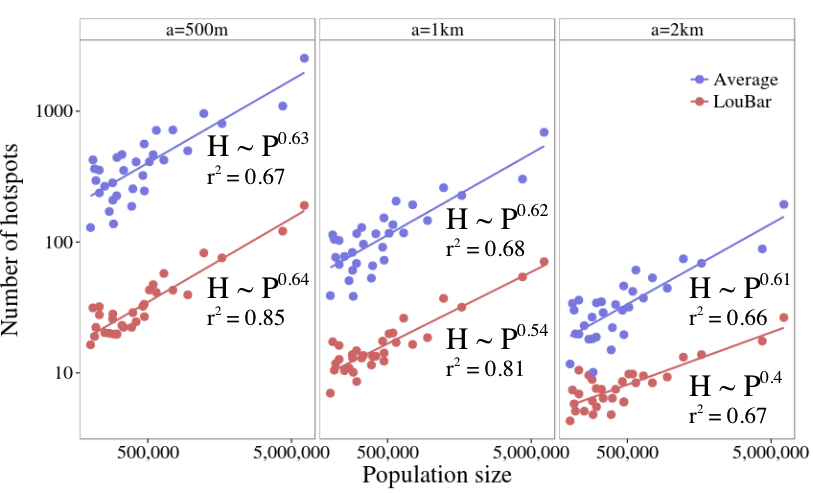}
\end{center}
\caption{{\bf Scatter plot and model fit line of the number of
    hotspots $H$ vs. the population size $P$ for the 31 cities
    studied.} Each point in the scatterplot corresponds to the average
  number of hotspots determined for each one-hour time bin of a
  weekday time period considered for the five weekdays. The linear
  relationship on a log-log plot indicates a power-law relationship
  between the two quantities, with an exponent value $\beta < 1$,
  indicating that the number of activity centers in a city grows
  sublinearly with its population size.}
\label{fig:HvsP2}
\end{figure}

In Figure~\ref{fig:HvsP2} we show the scaling relation between the
number of hotspots with the population and the effect of the grid
size. Here we see that the scaling results and the value of the
exponent are robust against a change in (i) the threshold used for
identifying the hotspots and (ii) the size of the grid cells.

\subsection*{Kendall's $\tau$}

\subsubsection*{Definition}

The Kendall rank coefficient is used as a test statistic to establish
whether two lists of random variables may be regarded as statistically
dependent. To each cell $i$ we associate its rank $r_i(t)$ in the
ordered density distribution at time $t$. Kendall’s $\tau$ value
indicates how much the hierarchy changed between $t-1$ and $t$. For a
set of pairs $(i,j)$, it is equal to the difference between the number
of converging pairs (i.e. $\rho_i$ was larger (resp. smaller) than
$\rho_j$ at $(t-1)$ and is still larger (resp. smaller) at $t$) and
the number of diverging pairs ($\rho_i$ was smaller (resp. larger)
than $\rho_j$ at $(t-1)$ and is larger (resp. smaller) at $t$). The
Kendall values $\tau(t)$ are plotted on Figure~\ref{fig:Kendall}.

Under the null hypothesis of independence of two lists, the
distribution of $\tau$ has an expected value of zero and for larger
samples, the variance is given by
\begin{equation}
\overline{\tau^2}=\frac{2(2n+5)}{9n (n-1)}
\end{equation}
Any value of $\tau$ larger than this null-value signals the existence
of relevant correlations.

\subsubsection*{Hierarchy stability}

\begin{figure*}[!ht]
\begin{center}
  \includegraphics[width=\linewidth]{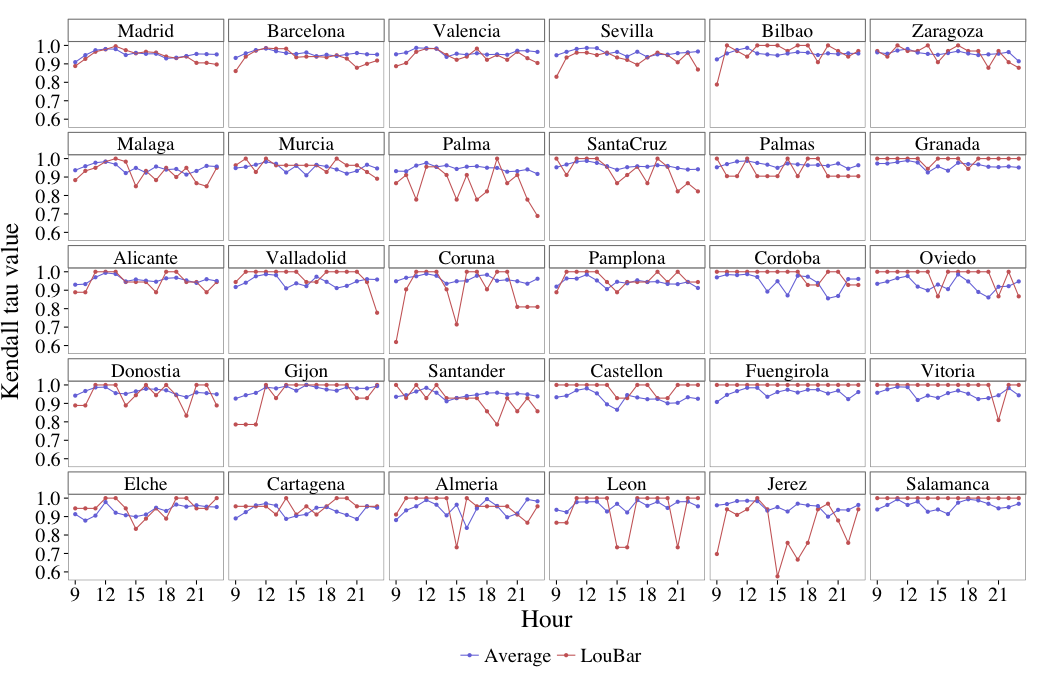}
\end{center}
\caption{{\bf Evolution of Kendall $\tau$ values for permament hotspots
    during daytime for an average weekday.}}
\label{fig:Kendall}
\end{figure*}

We show in Figure~\ref{fig:Kendall} the evolution of Kendall $\tau$
values calculated for the set of permament hotspots during daytime in
an average weekday, for 31 Spanish urban areas with more than 200,000
inhabitants. The curves are ranged by decreasing order of population
size (the biggest city in the top left corner, the smallest in the
bottom right). The red curves correspond to the daytime evolution of
the Kendall $\tau$ for the hotspots selected with the `LouBar' more
restrictive criterion, the blue ones to the Kendall $\tau$ of the
hotspots selected with the 'Average' criterion. These results indicate
that the hierarchy of permanent hotspots is indeed very stable in
time.

\section*{Acknowledgements}

The authors acknowledge funding from the EU commission through project
EUNOIA (FP7-DG.Connect-318367).

\section*{Author Contributions}

T.L. designed the study, analysed the data and wrote the manuscript;
M.L. processed and analysed the data; O.G.C and M.P. processed the
data; R.H. and J.J.R. coordinated the study; E.F.-M. obtained and
processed the data; M.B. coordinated and designed the study, and wrote the
manuscript. All authors read, commented and approved the final version
of the manuscript.

%\section*{References}
% The bibtex filename
%\bibliography{biblioPlos}

\end{document}